\documentclass[english]{article}
\usepackage{fullname}
\usepackage{fullpage}
\usepackage[ruled,vlined]{algorithm2e}
\usepackage{latexsym}
\usepackage{amsmath} \usepackage{amssymb}
\usepackage{amsfonts} \makeatletter
\usepackage{times}

\newcommand\secref[1]{\ifthenelse{\equal{#1}{}}{}{Section~\ref{s:#1}}}
\newcommand\inside{\beta}
\newcommand\reachfrom{B}
\newcommand\outside{\alpha}
\newcommand\reachto{A}
\newcommand\true{\algkey{true}}
\newcommand\false{\algkey{false}}

\newcommand{\alias}       [2]{\@ifdefinable #1{\let #1 #2}}
\alias\realias\let
\newcommand{\providealias}[2]{\@ifundefined #1{\let #1 #2}}
\newcommand{\mts}{MT_\Sigma}
\newcommand\ternary[3]{
  \left\{ \begin{array}{ll}
      {#2}
      & \textit{if }{#1}\\
      {#3}
      & \textit{otherwise}
    \end{array} \right.
}

  \newcommand\hastails{{tail nodes $T_{i}\subseteq V\equiv  T_{e_{i}}$}}
  \newcommand\hashead{{head $h_{i}\in V\equiv  h_{e_{i}}$}}

\newcommand\geneq{{\approx}}

\newcommand\comment[1]{}
\newcommand\into{\rightarrow}

\newcommand\logand{\wedge}
\newcommand\logor{\vee}
\newcommand\union{\cup}
\newcommand\intersect{\cap}
\newcommand\concat{\cdot}
\newcommand\bigconcat{\bullet}
\newcommand\assign{\leftarrow}
\newcommand\naturals{\mathbb{N}}
\newcommand\reals{\mathbb{R}}
\newcommand\positivereals{\mathbb{R}^{+}}

\newcommand\nonnegints{(\naturals \union\{0\})}
\newcommand\domain[1]{\text{dom }#1}
\alias\coll\mathcal
\newcommand\multiset[1]{\coll{M}(#1)}
\newcommand\seqn[2]{({#1}_{1},\ldots,{#1}_{#2})}

\newcommand\transformsto{\rightarrow}
\newcommand\derives{\Rightarrow}
\newcommand\logimplies{\implies} 
\newcommand\st{\;|\;}
\newcommand\bigst{\;\bigl | \;}
\newcommand{\nth}[2]{#1\textsuperscript{\textit{#2}}}
\newcommand{\kstar}{^{\star}}
\newcommand\algname[1]{\textbf{#1}}
\newcommand\algkey[1]{\algname{#1}}

\newcommand\cls[1]{\textbf{#1}}
\newcommand\url[1]{#1}
\newcommand\newVar[2]{\newcommand{#1}{#2}} 

\newcommand\PQ[1]{\algname{HEAP-{#1}}}
\newcommand\fnote[1]{({#1})}
\newcommand\Adj{\textit{Adj}}
\newcommand\Adji{\textit{Adj}^{-1}}

\newVar{\startnode}{\omega}

\newcommand\derivess[1]{\derives_{#1}}
\newcommand\derivesc[1]{\derives_{#1}\kstar}
\newcommand\derivesl[1]{\derives_{#1}^{L*}}

\newcommand\algref[1]{Algorithm~\ref{#1}}

\title{Context-free Algorithms}   
\author{Jonathan Graehl \\ SDL Research \thanks{Work done at University of Southern California, Information Sciences Institute, 4676 Admiralty Way, Marina del Rey, CA 90292}
}
\date{July 20, 2005}    

\begin{document}
\maketitle

\begin{abstract}
  Algorithms on grammars/transducers with context-free derivations: hypergraph reachability, shortest path, and inside-outside pruning of 'relatively useless' arcs that are unused by any near-shortest paths.

\end{abstract}

\section{Introduction}
We present algorithms on context-free grammars (and also on hypergraphs and regular tree grammars, which share the same context-free derivation rule): hypergraph reachability, shortest path, and inside-outside pruning of 'relatively useless' arcs that are unused by any near-shortest paths. \secref{notation} is optional for those already familiar with regular tree grammars (analogous to derivation trees of context free grammars) and/or hypergraphs.

\section{Notation}
\label{s:notation}
\label{sec1}
\subsection{Strings}

$\Sigma\kstar$ are the \emph{strings over alphabet $\Sigma$}.  For
$s=(s_{1},\ldots,s_{n})$ the \emph{length} of $s$ is $|s|\equiv n$ and the
$i$\emph{th letter} is $s[i]\equiv s_{i}$, for all $i\in indices_{s}\equiv\{
i\in\naturals \st 1\leq i\leq n\}$, and the concatenation of a sequence of
letters by index is $s[\seqn{f}{n}\in indices_{s}\kstar]\equiv
(s[f[1]],\ldots,s[f[n]])$.  \emph{Concatenation} of strings is specified by
the $\concat$ operator, where $a\concat b\equiv(a[1],\ldots,a[|a|],b[1],\ldots,b[|b|])$.
\comment{Naturally, $|a\concat b|=|a|+|b|$.}
\comment{:
\[
\seqn{a}{n}\concat\seqn{b}{m}\equiv(a_{1},\ldots,a_{n},b_{1},\ldots,b_{m})\]
}

\comment{
The \emph{letters in $s$} are $letters_{s}=\{
l|\exists i\in{indices_{s}}:s[i]=l\}$.
The \emph{spans} of $s$ are
$spans_{s}=\{(a,b)\in{\{\mathbb{{N}}^{2}\st1\leq a\leq b\leq n+1\}}$, and the
\emph{substring at span $p=(a,b)$} of $s$ is $s\downarrow p\equiv(s_{a},\ldots
s_{b-1})$, with $s\downarrow(a,a)=()$.
The \emph{subsequences} of $s$ are given
by a \emph{subsequence map} $f\in subseqmap_s$:
\[
subseqmap_s\equiv
\{\seqn{i}{n}\in indices_s\kstar\st i_1 < \ldots <  i_n \}
\]
A subsequence of $s$ by map $f$ is $s[f]$.  (The subsequences of $s$ are $subseq_s\equiv \{s[f]\st f\in
subseqmap_s\}$).  For a letter $\sigma\in \Sigma$ there is exactly one maximal
subsequence consisting of repetitions of that letter, and its map is
$subseqmap_s(\sigma)$:
\[
subseqmap_{s}^{\geneq}(\sigma)\equiv \bigconcat_{i=1}^{|s|} \ternary{s[i]\geneq\sigma}{(i)}{()}
\]
Note the $\geneq$ superscript, which, if omitted, is assumed to be the usual
equality ($=$).  Different $\geneq$ predicates can be useful for matching on
projections of $\Sigma$.  This convention will be assumed throughout.

We can extend a function $f:\Sigma\into \Delta$, to sequences by mapping it over each element $f:\Sigma\kstar\into\Delta\kstar$, where $f(s\in\Sigma\kstar)=(f(s[1]),\ldots,f(s[|s|]))$.
}

\subsection{Multisets}

A \emph{multiset $M$ of $S$} is a partial function $M:S\into \naturals$, or
equivalently, a functional binary relation $M\subset S\times \naturals$.  The
class of multisets of $S$ is written $\multiset{S}$.  If $M(s)=m\in{\naturals}$,
we say $(x,m)\in{M}$, $x\in{M}$, and the \emph{multiplicity of $x$ in $M$} is
$m$.  Intuitively, the multiplicity is the number of times an element occurs.
The \emph{domain of $M$} is $\domain{M}\equiv \{x\in{M}\}$. In some cases it is
convenient to interpret $M$ as a total function from $S\rightarrow \nonnegints$
where $M(x\notin{\domain{M}})\equiv 0$.  A set $S$ can be interpreted as a
multiset where each $x\in{S}$ has multiplicity $S(x)\equiv 1$.  A sequence
$V=\seqn{v}{n}\in{S\kstar}$ can also be seen as a multiset with
$V(x)\equiv \sum_{i:v_{i}=x}1$ (after all, another notation of a multiset is just a
set listed without removal of duplicates, e.g. $\{a,b,a\}$).

\comment{
The \emph{intersection}, or \emph{product}, of multisets $M$ and $N$ is $M\intersect N \equiv \{(x,ab)\st
(x,a)\in{M} \logand (x,b)\in{N}\}$.  Their \emph{union}, or \emph{sum} is $M\union N$ defined by
$M\union N:(\domain{M})\union (\domain{N}) \into \naturals$ where $(M\union
N)(x)\equiv M(x)+N(x)$.  The \emph{size} of a multiset $M$ is $|M|\equiv
\sum_{x\in{M}}M(x)$.  A multiset $M$ can be \emph{scaled} by a constant $k\in
\naturals$: $kM\equiv \{(x,km)\st (x,m)\in M\}$.

The \emph{factorial of a multiset $M$} is the set of unique permutations
$M!\subset \Sigma\kstar$ that are equivalent to $M$ when considered as a
multiset.  The number of unique permutations of a multiset $M$ is given by
\[
|M!|=\frac{|M|!}{\prod_{(x,m)\in{M}}m!}
\]
since all the $M(x)!$ ways of reordering
the $M(x)$ identical items $x\in M$ are indistinguishable.  The multiset
factorial of a sequence can be generated in the tradition of sequence
permutations, except doing nothing when two items to be swapped are equal,
instead of explicitly counting the multiplicity of the unique elements.
}

\subsection{Trees}

$T_{\Sigma}$ is the set of \emph{(rooted, ordered, labeled, finite) trees over
  alphabet $\Sigma$.}

$T_{\Sigma}(X)$ are the \emph{trees over alphabet $\Sigma$, indexed by $X$}---the subset of $T_{\Sigma\union X}$ where only leaves may be labeled by $X$.
($T_{\Sigma}(\emptyset)=T_{\Sigma}$.)  \emph{Leaves} are nodes with no children.

The \emph{nodes} of a tree \emph{t} are identified one-to-one with its
\emph{paths}: \emph{}$paths_{t}\subset paths\equiv \naturals \kstar \equiv\bigcup_{i=0}^{\infty}\naturals^{i}$ ($A^{0}\equiv\{()\}$).  The path
to the root is the empty sequence $(),$ and $p_{1}$ \emph{extended by} $p_{2}$
is $p_{1}\concat p_{2}$, where $\concat$ is concatenation.

For $p\in{paths_{t}}$, $rank_{t}(p)$ is the number of children, or \emph{rank},
of the node at $p$ in $t$, and $label_{t}(p)\in{\Sigma} \union X$ is its \emph{label}. The \emph{root of $t$} is $root(t)=label_{t}(())$. The
\emph{ranked label} of a node is the pair
$labelandrank_{t}(p)\equiv(label_{t}(p),rank_{t}(p))$. For $1\leq i\leq
rank_{t}(p)$, the \nth{$i$}{th} \emph{child} of the node at $p$ is located at
\emph{path} $p\concat(i)$.  The \emph{subtree at path $p$ of $t$} is $t\downarrow p$,
defined by $paths_{t\downarrow p}\equiv\{ q\st p\concat q\in{paths_{t}}\}$ and
$labelandrank_{t\downarrow p}(q)\equiv labelandrank_{t}(p\concat q)$.  The
\emph{children of $t$} are $children_t\in T_\Sigma\kstar$, with
$children_t[i]=t\downarrow (i), \forall 1\leq i \leq rank(t)$.

The \emph{paths to $X$ in $t$} are $paths_{t}(X)\equiv\{ p\in{paths_{t}}\st label_{t}(p)\in{X}\}$.
A \emph{frontier} is a set of paths $f$ that are \emph{pairwise
  prefix-independent}:
\[
\forall p_{1},p_{2}\in{f},p\in{paths}:p_{1}=p_{2}\concat p\logimplies
p_{1}=p_{2}\]

A \emph{frontier of t} is a frontier $f\subseteq paths_{t}$.

For $t,s\in{T_{\Sigma}(X)},p\in{paths_{t}}$, $t[p\assign s]$ is the
\emph{substitution of $s$ for $p$} in $t$, where the subtree at path $p$ is
replaced by $s$. For a frontier $f$ of $t$, the \emph{mass substitution of $X$
  for the frontier $f$ in $t$} is written $t[p\assign X,\forall p\in{f}]$ and
is equivalent to substituting the $X(p)$ for the $p$ serially in any order.

The \emph{yield of $X$ in} $t$ is $yield_{t}(X)$,
\comment{
  the concatenation (in lexicographic order\footnote{$()<_{lex}(a)$, $(a_{1})<_{lex}(a_{2}) \textrm{ iff } a_{1}<a_{2}$,  $(a_{1}) \cdot b_{1}<_{lex} (a_{2})\cdot b_{2} \textrm{ iff } a_{1}<a_{2} \logor (a_{1}=a_{2} \logand b_{1}<_{lex} b_{2})$}) over paths
  to leaves $l\in{paths_{t}}$ (such that $rank_{t}(l)=0$) of
  $label_{t}(l)\in{X}$---that is,
}
the string formed by reading out the leaves
labeled with $X$ in left-to-right order.
The usual case (the \emph{yield of
  $t$}) is $yield_{t}\equiv yield_{t}(\Sigma)$.

We may also consider the \emph{monadic strings} in $t$, $mstrings_t \subset \Sigma\kstar$, obtained by reading off the
labels along some path from the root down.
The paths that read off a monadic string $s$ in $t$ are $mpaths_{t}^{\geneq}(s)\equiv \{p\in
paths_t\st \forall 1\leq i \leq |p|+1 : label_t(p\downarrow (1,i))\geneq s[i]\}$, and the string of labels along a path is
$mstring_t(p\in paths_t)\equiv \bigconcat_{i=1}^{|p|+1} (label_t(p\downarrow
(1,i)))$ (so $\forall p\in mpaths_t^{\geneq}(s) : mstring_t(p)\geneq s$).  Then $mstrings_t \equiv \{mstring_t(p\in paths_t)\}$ and
$t\downarrow s$ is the sequence of \emph{subtrees of $t$ along the monadic
  string $s$} (in lexicographic path order):
\[
t\downarrow^{\geneq} s\in{mstrings_t} \equiv \bigconcat_{p\in mpaths_{t}^{\geneq}(s) \text{ in
    lexicographic order }} (t\downarrow p)
\]
Naturally, the path in $t$ to the \nth{$i$}{th} element of $t \downarrow s$ is
the \nth{$i$}{th} (in lexicographic order) $mpaths_t(s)$.

\comment{
The \emph{$l$-labeled-children of $t$} are contained in the subsequence
$children_{t}^\geneq(l)=t\downarrow^\geneq (l)$:
\[
children_t^\geneq(l)\equiv  {
  c[subseqmap_{c,=_r}(l)] \text{ where } c\equiv children_t \text{ and } a=_rb \text{ iff } root(a)\geneq b
}
\]
}

\subsection{Regular Tree Grammars}

A \emph{weighted regular tree grammar} (\cls{wRTG}) $G$ is a quadruple
  $(\Sigma,N,S ,P)$, where $\Sigma$ is the alphabet, $N$ is the finite set of
  \emph{nonterminals}, $S \in{N}$ is the \emph{start (or initial) nonterminal},
  and \emph{$P\subseteq N\times T_{\Sigma}(N)\times\positivereals $} is the
  finite set of \emph{weighted productions} ($\positivereals \equiv\{
  r\in{\reals }\st r>0\}$).
  We define the binary relation $\derivess{G}$ (\emph{single-step derives in G})
  on $T_{\Sigma}(N) \times (paths \times P)\kstar$, pairs of trees and \emph{derivation histories}, which are logs of (location, production used):
  \[
  \begin{array}{r}
    \derives_{G}\equiv\Bigl\{((a,h),(b,h\concat (p,(l,r,w)))\bigst \\
    (l,r,w)\in{P}\logand p\in{paths_{a}(\{ l\})}\logand b=a[p\assign r]\Bigl\}
  \end{array}
  \]

  where $(a,h)\derivess{G}(b,h\concat (p,(l,r,w)))$ iff tree $b$ may be derived
  from tree $a$ by using the rule $l\transformsto^{w}r$ to replace the nonterminal
  leaf $l$ at path $p$ with $r$.  For a derivation history
  $h=((p_{1},(l_{1},r_{1},w_{1})),\ldots,(p_{n},(l_{1},r_{1},w_{1})))$, the \emph{weight of $h$} is $w(h) \equiv \prod_{i=1}^{n} w_{i}$, and call $h$ \emph{leftmost} if $L(h)\equiv \forall 1\leq i < n :  p_{i+1} \nless_{lex} p_{i}$.\footnote{$()<_{lex}(a)$, $(a_{1})<_{lex}(a_{2}) \textrm{ iff } a_{1}<a_{2}$,  $(a_{1}) \cdot b_{1}<_{lex} (a_{2})\cdot b_{2} \textrm{ iff } a_{1}<a_{2} \logor (a_{1}=a_{2} \logand b_{1}<_{lex} b_{2})$}

  The reflexive, transitive closure of
  $\derivess{G}$ is written $\derivesc{G}$ (\emph{derives in $G$}), and
  the restriction of $\derivesc{G}$ to leftmost derivation histories is
  $\derivesl{G}$ (\emph{leftmost derives in $G$}).

  The \emph{weight of $a$ becoming $b$ in $G$} is $w_{G}(a,b) \equiv
  \sum_{h:(a,())\derivesl{G}(b,h)}w(h)$, the sum of weights of all unique
  (leftmost) derivations transforming $a$ to $b$, and the \emph{weight of $t$ in $G$} is
  $W_{G}(t)=w_{G}(S ,t)$.  The \emph{weighted regular tree language produced by
    $G$} is $L_{G}\equiv \{(t,w)\in T_{\Sigma} \times \positivereals \st
  W_{G}(t)=w \}$.

  The \emph{derivation tree grammar} for a \cls{wRTG} $G=(\Sigma,N,S,P)$ is $DG(G)=(P,N,S,P')$, where
  \[
  P'\equiv\{(l,p(yield_{N}(r)),w)\st p=(l,r,w)\in{P}\}
  \]
  ($p((s_{1},\ldots,s_{n})\in N\kstar)$ is the tree with root label $p$, rank $n$, and \nth{i}{th} child leaf $s_{i}$).  The produced trees are called \emph{derivation trees} and correspond one-to-one with tree-producing derivations in $G$.


\comment{
  \subsection{Unordered Trees}

  Just like a multiset is a sequence where we don't care about the order of its
  elements, we can consider trees where we don't care about the order of
  children. Call $\mts=\{t=(root(t)\in \Sigma,children_t\in \multiset{\mts})\}$ the set of \emph{(rooted, labeled, finite) unordered trees
    over alphabet $\Sigma$}.

  The root's label is \emph{$root(t)$}; its rank would be
  $|children_t|$).  Paths are not defined for unordered trees, but, as
  with ordered trees, we are interested in the subtrees descended along paths
  labeled by a monadic string, $t\downarrow s$:
  \[
  t\downarrow^\geneq \seqn{s}{n}\equiv \ternary{n>1}{
    \union_{(c,m)\in{t\downarrow (s_1)}}m(c\downarrow (s_2,\ldots,s_n))
  }{\ternary{root(t)\geneq s_1}{\{t\}}{\emptyset}}
  \]
  As for ordered trees, we let $mstrings_t$ be a multiset (instead of a sequence) with
  multiplicity for $s$ $mstrings_t(s)\equiv |t\downarrow s|$.
\comment{
Similarly, we define $children_{t}^{\geneq}(l)\equiv t\downarrow^{\geneq} (l)$.
}

  An ordered tree $t$ can be interpreted as an unordered tree $u$ by the recursive
  rule $U$ that $U(t)\equiv (root(u)\equiv root(t), children_u\equiv U(children_t)$
  (sequence interpreted as a multiset).  Then, properties related to monadic
  strings $s$ of $t$ should be the same multiset in the ordered $t$ as in
  $u$---for example, $t \downarrow s=u \downarrow s$.
}

  \subsection{Hypergraphs}

  A \emph{(directed) hypergraph $G$} is a pair $G=(V,E)$ where $V$ is a set
  of \emph{vertices} (or \emph{nodes}) of $G$, and $E$ are the \emph{edges} (or
  \emph{hyperarcs}) of $G$.  An edge
  $e=(h_{e}\in{V},T_{e},c_{e}:\reals^{|T_{e}|}\into \reals)$ has \emph{head
    $h_{e}$}, \emph{tails $T_{e}$}, and \emph{cost function $c_{e}$}.  The cost
  function for an edge maps the costs of reaching its tails to the cost of
  reaching the head through that edge.

  In a hypergraph, $T_{e}\subseteq V$---the tails are subsets of the vertices.

\comment{For a \emph{multi-hypergraph}, $T_{e}\in{\multiset{V}}$---the tails are a
  multiset of vertices.  } In an \emph{ordered multi-hypergraph},
  $T_{e}\in{V\kstar}$---the tails are ordered sequences.

  Typically hyperarc cost functions are symmetric; if not, then the order of
  arguments is the same as the order of tails.
, or for unordered hypergraphs,  fixed by some arbitrary total order $<_{G}$ on $V$.
  The usual cost function
  is given by $c_{e}\seqn{x}{n}\equiv l_{e}+\sum_{i=1}^{n}x_{i}$, where $l_{e}$
  is the \emph{length} of the edge.  A typical asymmetric cost function would
  combine tail hyperpath costs with different weights for each tail.

  We say there is a \emph{hyperpath from $X\subseteq V$ to $y\in{V}$ in
  $G=(V,E)$}, written $X\leadsto_{G}y$, if $y\in{X}\logor \exists e\in{E} :
  h_{e}=y \logand \forall t\in{T_{e}} : X\leadsto_{G}t$.  A \emph{hyperpath-tree
  $t\in{(X\leadsto_{G}y)}$} is a tree labeled by edges, corresponding to a proof
  of $X\leadsto_{G}y$ (with a separate proof for each multiple occurrence of a
  tail vertex - note: the usual B-hyperpath allows only a single incoming
  hyperarc/proof of each vertex - our hyperpath-trees are more like derivations
  in a context-free grammar).  The \emph{cost} of a hyperpath-tree $p$ is
  written $c(p)$ and is computed bottom-up for each subtree with root label $e$
  using $c_{e}$.

  For any derivation grammar $G'=(P,N,S,P')$ of \cls{wRTG} $G=(\Sigma,N,S,P)$,
  there is an equivalent ordered multi-hypergraph $H=(N\union\{\startnode\},E)$ with an edge $e\in{E}$
  for each production $p=(l,r,w)\in{P'}$ such that $h_{e}=l$,
  $T_{e}=\ternary{yield_{N}(r)=\emptyset}{\{\startnode\}}{yield_{N}(r)}$, and the usual
  cost function with $l_{e}=-\ln{w}$.  The hyperpath-trees $\startnode \leadsto_{H} S$ are exactly
  the derivation trees for $G$, with the cost of the hyperpath-tree equal to the $\ln$
  of the weight of the tree (obviously, the labels of the hyperpath-tree are
  $e\in{E}$ and the labels of the derivation tree are $p\in{P}$, but there is an
  isomorphism between them, due to the construction of $E$).

  A hypergraph $(V,E)$ may be interpreted as a multigraph $(V,E')$ with an edge for every
  tail of each hyperarc ($E'=\{(h_e,t\in T_e,c_e)\st (h_e,T_e,c_e)\in E\}$).  We can
  refer to \emph{simple} (or \emph{monadic}) paths corresponding to the usual
  paths in the graph.  In fact, monadic strings $s$ of hyperarcs from a hyperpath-tree
  for $(V,E)$ correspond to a simple path in $h_{s[|s|]}\leadsto_{(V,E')}h_{s[1]}$.

  \section{Pruning Along a Hyperpath-Tree}

  If we are only interested in hyperpath-trees $X\leadsto_G y$, we can \emph{prune
    $G$ along $X$ to $y$} by eliminating vertices and hyperarcs that don't
  appear in any (cheap) hyperpath-tree.  This is analogous to the problem of reducing
  a context free grammar by eliminating useless nonterminals \cite{hopcroft},
  except that we wish to also eliminate those useful only for high-cost
  hyperpath-trees.

  Since we care only for the existence of a (cheapest) path for each node, tails
  of edges may be considered as sets while addressing this problem, so that
  multiply appearing tails $t$ in a multi-hypergraph always reuse the same
  hyperpath-tree $X\leadsto_G t$.  We assume the cost function $c_e(c)=l_e+\sum_{(t,m)\in
    T_e}w_e(t)mc(t)$, where $c(t)$ is the cost due to the hyperpath-tree $X\leadsto
  t$ and $w_e(t)$ is a weight given to $t$-tails of that edge.

  Unweighted pruning consists of first eliminating vertices (and hyperarcs they
  occur in) that cannot be reached from the start, and second, eliminating from
  the remainder all those that do not lie along any hyperpath-tree to the destination.
  The first step can be performed in linear time by \algref{algo_reachfrom}.

  \begin{algorithm}
    \DontPrintSemicolon

    \caption{
      Single-source-set hypergraph reachability
    }

    \KwIn{

      A set of source nodes $X\subseteq V$ in a hypergraph $G=(V,E)$, nodes $V$,
      and hyperarcs $E=\{e_1,\ldots,e_m\}$ indexed by $1\leq i\leq m$.  Each hyperarc
      has \hastails and \hashead.

    }

    \KwOut{

      For all $y\in{V}$, $\reachfrom[y]=\true$ if $X\leadsto_G y$, $\false$ otherwise.
      Time complexity is $O(t)$ where $t$ is the total size of the input.

    }

    \Begin{

      \lFor{$y\in{V}$}{
        $\reachfrom[y] \assign \false$\;
        $\Adj[y] \assign \{\}$\;
      }

      \For{$1\leq i \leq e$,\text{ index of a hyperarc }$(T_{i}=\{x_{1},\ldots,x_{k}\}) \rightarrow \{h_{i}\}$}{
        $r[i] \assign k$\;
        \tcc{ $r[i]$ is the number of tail nodes remaining before edge $i$ fires.}
        \lFor{$1\leq j \leq k$}{$\Adj[x_{j}] \assign \Adj[x_{j}] \union \{i\} $\;
        }
      }

      \lFor{$y\in X$}{\algname{REACH}(y)\;}

    }
    \BlankLine
    $\algname{REACH}(y)\equiv$
    \Begin{
      \If{$\neg \reachfrom[y]$}{
        $\reachfrom[y]\assign \true$\;
        \For{$i\in{\Adj[y]}$}{
          \If{$\neg \reachfrom[h_i]$}{
            $r[i] \assign r[i] - 1$\;
            \lIf{$r[i] = 0$}{$\algname{REACH}(h_i)$\;}
          }
        }
      }
    }
    \label{algo_reachfrom}
  \end{algorithm}

  The weighted version of \algref{algo_reachfrom} establishes the
  lowest cost way of reaching each vertex from a start set (or that
  there is none).  \algref{algo_knuth}, adapted from
  \cite{knuthgrammar} (first published in \cite{poweroftree}), is an
  extension of the graph shortest path problem \cite{dijkstra} to the
  hypergraph case.  It works the same except that vertices are visited
  in increasing order of the cost of reaching them from $X$, and so
  requires a priority queue.  Activated hyperarcs serve to potentially
  lower the cost of reaching their head, but visiting the head is
  deferred until it is certain that its minimal cost hyperpath-tree is
  known.  This is in contrast to the simple depth first approach in
  the unweighted case, where the head is visited immediately with a
  recursive function call (using the implicit program stack for
  queuing nodes).

\newcommand\sink{\omega}
\newcommand\countnonterm{{\#}}

  \begin{algorithm}
    \DontPrintSemicolon

    \caption{
      \algname{ViterbiInside}: single-source-set, multi-destination shortest hyperpath-trees.}

    \KwIn{

      A set of source nodes $X\subseteq V$ with initial costs $\{i_{x},\forall
      x\in{X}\}$, and a hypergraph with $n$ nodes $V$, and $m$ hyperarcs
      $\seqn{e}{m}$ indexed by $1\leq i\leq m$.  Each hyperarc has \hastails, \hashead, and superior cost
      function $c_{i}\equiv c_{e_{i}}$ \fnote{$f$ is \emph{superior} iff
        $f(x_{1},\ldots,x_{k}) \geq x_{i}, \forall 1\leq i\leq k$
        \cite{knuthgrammar}} of variables $T_{i}$.  The cost functions are
      implemented by constant time operations \algname{BIND}($c_{i},y\in
      T_{i},\text{cost of }y$) and \algname{INF}($c_{i}$), which returns a lower
      bound on the cost given the variables bound so far.

      For a context-free grammar or regular tree grammar, introduce a fictitious
      sink nonterminal $\sink$ to the rhs of terminal rules.  Now let the $V$ be
      the nonterminals, and let $X$ be ${\omega}$.  For each \nth{i}{th} rule,
      let $h_{i}$ be the lhs nonterminal, $T_{i}$ be the set of rhs nonterminals
      (or ${\sink}$ if there are none).  Finally, initialize
      \algname{INF}($c_{i}$) to $w_{i}=-\log{P(i|h_i)}$, the negative log rule
      probability of rule $i$, and define \algname{BIND}($c_{i},y\in T_{i},c$)
      as increasing \algname{INF}($c_{i}$) by $\countnonterm_{i}(y)c$, where
      $\countnonterm_{i}(t)$ is the number of occurrences of nonterminal $t$ in
      rule $i$.

    }

    \KwOut{

      For all $v\in{V}$, $\pi[v]=i$ is the index of the cheapest hyperarc with
      head $h_{i}=v$, giving the predecessor
      relation of the cheapest unordered hyperpath-tree from the $X \leadsto t$), and
      $\inside[v]$ is minimum cost of reaching $v$.  $\pi[v]=0$ if there is no
      cost-improving edge to $v$.  Time complexity is $O(n\lg{n}+t)$ where ($t$
      is the total size of the input) if a Fibonacci heap is used, or
      $O(m\lg{n}+t)$ if a binary heap is used.

    }

    \Begin{

      \For{$y\in{V}$}{
        \lIf{$y\in{X}$}{$\inside[y] \assign i_{y}$\;}
        \lElse{$\inside[y] \assign \infty$\;}
        $\pi[y] \assign 0$\;
        $\Adj[y] \assign \{\}$\;
      }

      $Q \assign \PQ{CREATE}()$\;

      \lFor{$x\in{X}$}{$\PQ{INSERT}(Q,x,i_{x})$\;}

      \For{$1\leq i \leq m$,\text{ index of a hyperarc }$(T_{i}=\{x_{1},\ldots,x_{k}\}) \rightarrow^{c_{i}} \{h_{i}\}$}{
        $r[i] \assign k$\;
        \tcc{ $r[i]$ is the number of tail nodes remaining before edge $i$ fires.}
        \lFor{$1\leq j \leq k$}{$\Adj[x_{j}] \assign \Adj[x_{j}] \union \{i\} $\;
        }
      }

      \While{$Q \neq \emptyset$}{
        $y \assign \PQ{EXTRACT-MIN}(Q)$\;
        \For{$i\in{\Adj[y]}$}{\tcc{ edge $i$ with $y$ as a tail}
          \If{$\algname{INF}(c_{i}) < \inside[h_{i}]$}{
            $\algname{BIND}(c_{i},y,\inside[y])$\;
            $r[i] \assign r[i] - 1$\;
            \If{$r[i] = 0$}{
              $c \assign \algname{INF}(c_{i})$\;
              \If{$c < \inside[h_{i}]$}{
                \lIf{$\inside[h_{i}] = \infty$}{$\PQ{INSERT}(Q,h_i,c)$\;}
                \lElse{$\PQ{DECREASE-KEY}(Q,h_{i},c)$\;}
                $\pi[h_{i}] \assign i$\;
                $\inside[h_{i}] \assign c$\;
              }
            }
          }
        }
      }

    }

    \label{algo_knuth}
  \end{algorithm}

  Having eliminated parts of the hypergraph that aren't reachable from $X$, it still remains to further remove any parts that don't contribute to reaching $y$.  In \algref{algo_reachto}, we perform a simple depth-first traversal from heads to tails of
  hyperarcs, starting with the destination $y$, ultimately
  saving only vertices that can help reach $y$.

  \newcommand\hrestrict[2]{{{#1}\langle {#2} \rangle}}

  To see how this works, let the \emph{restriction} of hypergraph $G=(V,E)$ to a subset of
  its vertices $V'\subseteq V$ be $\hrestrict{G}{V'}\equiv(V',E):E'=\{e\in E \st
  h_e\in V' \logand T_e\subseteq V'\}$.  First, run \algref{algo_reachfrom} on
  $G$ to find $V'=\{v \in V' \st X\leadsto_G\}$, then second, run
  \algref{algo_reachto} on the resulting restriction $G'=\hrestrict{G}{V'}$ to
  find $V''=\{v\in V' \st \exists F\supseteq\{v\}: F \leadsto_{G'} y$.  Then the
  hypergraph $G''=\hrestrict{G'}{V''}$ has the same hyperpath-trees $X\leadsto_{G''}
  y$ as $G$, and is the minimal such.

  The order of these steps is essential - there
  may be vertices that only help reach $y$ through hyperarcs that are eliminated
  in \algref{algo_reachfrom}.  In the second step, we qualify each node $t\in
  T_e$ that is connected through $e$ to $y$ as
  participating in a path to $X \leadsto_G h_e$ automatically, which is sound only if we can assume some
  path from $X \leadsto_G t'$, for all $t'\in T_e$.  But the first step guarantees this by removing all nodes that aren't reachable from $X$.

  \begin{algorithm}
    \DontPrintSemicolon

    \caption{
      Single-destination hypergraph reachability
    }

    \KwIn{

      A destination node $y\in V$ in a hypergraph $G=(V,E)$, with $n$ nodes $V$,
      and $m$ hyperarcs $E=\{e_1,\ldots,e_m\}$ indexed by $1\leq i\leq m$.  Each hyperarc
      has \hastails and \hashead.

    }

    \KwOut{

      For all $x\in{V}$, $\reachto[x]=\true$ if there is a hyperpath-tree
      $X\leadsto_G y$ such that $x\in{X}$, $\false$ otherwise.  Time complexity
      is $O(t)$ where $t$ is the total size of the input (this is simple
      depth-first search on the projected regular graph).

    }

    \Begin{
      \lFor{$x\in{V}$}{
        $\reachto[x] \assign \false$\;
      }
      \algname{USE}(y)\;
    }
    \BlankLine
    $\algname{USE}(y)\equiv$
    \Begin{
      $\reachto[y]\assign \true$\;
      \For{$t\in T_i$}{
        \If{$\neg \reachto[t]$}{
          $\algname{USE}(t)$\;
        }
      }

    }
    \label{algo_reachto}
  \end{algorithm}

  What we are really doing is reversing a hypergraph by interpreting it as a
  monadic graph consisting of all edges formed by selecting just one tail of
  each hyperarc, and plugging in a default rule for completing the omitted
  siblings.  We can extend this strategy to the weighted case, using the
  shortest hyperpath-tree $X\leadsto v$ ($\pi[v]$) (from from
  \algref{algo_knuth}) for each omitted sibling $v$.  Then we can attribute to
  each monadic arc the cost of those omitted hyperpath-trees ($\inside[v]$), in
  addition to the cost of its original hyperarc.  Then we can perform the usual
  single-source shortest graph paths computation\cite{dijkstra} on the this
  reverse monadic graph.

  Since any subtree of a shortest hyperpath-tree $t\in{(X\leadsto y)}$ is a
  shortest hyperpath-tree from $X$ to its root-head $h_{label_{t}(())}$, we can
  decompose the shortest hyperpath-tree using node $v$ into the shortest
  \emph{inside} $X\leadsto v$ plus the \emph{outside} $v\leadsto y$ formed by
  reconstituting a path in the monadic graph with the default interpretation of
  omitted siblings.  The outside part is an almost-hyperpath-tree, missing only
  an inside subtree for $X\leadsto v$ (an outside tree would be a hyperpath-tree
  from $X\union \{v\} \leadsto y$).  This is the insight behind the
  inside-outside algorithm\cite{InsideOutside} for training context free string
  grammars, and also its extension to training tree transducers\cite{TTT}.

  Note that this decomposition means that the cost functions for hyperarcs must
  be separable into an independent sum over parts due to the tails and a part
  due to the arc.

  In \algref{algo_dijkstra}, we implicitly perform this reversal and
  monadification of a hypergraph and obtain for each vertex $v$ the cheapest way
  to complete the hyperpath-tree $X\leadsto v$ into $X\leadsto v\leadsto y$ (by
  that we mean adjoining some inside hyperpath-tree $X\leadsto v$ with , using
  parent $\psi[v]$ with total outside cost (leaving out the cost of $X\leadsto
  v$) $\outside[v]$.

  Then, the \emph{utility} of $v$, or the cost of the
  cheapest hyperpath-tree using it, is just $\gamma[v]\equiv
  \outside[v]+\inside[v]$ and the utility of hyperarc $e$ is $\gamma[e]\equiv
  \outside[h_e]+l_e+\sum_{(t,m)\in T_e}m\inside[t]$.  It is then easy to select
  vertices and edges for removal based on some criteria on their utility
  relative to the cost of the cheapest hyperpath-tree $X\leadsto y$, which is
  $\inside[y]$.

  \algref{algo_prune_relatively_useless} selects the minimal
  subset of the hyperarcs and vertices necessary to include the best
  hyperpath-tree $x\leadsto y$ with cost $\inside[y]$ and all hyperpath-trees
  with cost no worse than $\inside[y]+\delta$.

\newcommand\holdout[3]{{\algname{COSTEXCEPT_{#1}({#2},{#3})}}}

  \begin{algorithm}
    \DontPrintSemicolon

    \caption{
      \algname{ViterbiOutside} - single-destination, shortest outside hyperpath-trees
    }

    \KwIn{

      A destination $y\in V$ and default (inside) costs $\inside[v]$ for
      reaching each $v\in V$ from $X$ (computed with \algname{ViterbiInside}), for a
      hypergraph with $n$ nodes $V$, and $m$ hyperarcs $\seqn{e}{m}$ indexed by
      $1\leq i\leq m$.

\comment{
      Each hyperarc has \hastails, \hashead, and superior cost function
      $c_{i}\equiv c_{e_{i}}$.  The cost function is provided as an amortized constant time operation that builds up the cost of using the default cost ways to reach the tails of an edge, then taking the edge, but holding out one instance of a tail $v$,
      $\holdout{\inside}{i}{v\in V}$, for example, in $\holdout{\inside}{i}{v}\equiv (l_{e_i}+\sum_{(t,m)\in
        T_{e_i}}m\inside[t]) - \inside[v]$, everything but the last term (a constant time operation) is constant with respect to v and the constant takes just O($|\domain(T_e)|$) time to compute.
}
      Each hyperarc has length (i.e. cost to use) $l_i\equiv l_{e_i}$, a
      multiset of tails $T_i\equiv T_{e_i}\in \multiset{V}$, and \hashead.  The cost for
      hyperpath-tree from $X\leadsto h_e$ using edge $e$ and the best hyperpath-trees from $X$ to each of its tails $t$ with cost $\inside[t]$ is
      $c_{e}=l_{e}+\sum_{(t,m)\in T_{i}}m\inside[t]$ (where m is the number of occurrences of $t$ in the tails), but other cost functions are possible - what is
      important is the ability to build up the cost for using an edge assuming the default for its tails, and later subtract out the contribution from the default of a single
      instance of a tails.

  }

    \KwOut{

      For all $v\in V$, $\psi[v]$ is the index of the hyperarc used to reach
      $y$ from $v$ (or 0 if none was taken) with the minimum outside cost $\outside[v]$=$\inside[y]-\inside[v]$ given by assuming
      the default cost way to was used to reach its siblings from $X$.  Time complexity is
      $O(n\lg{n}+t)$ where ($t$ is the total size of the input) if a Fibonacci
      heap is used, or $O(m\lg{n}+t)$ if a binary heap is used.

    }
    \Begin{
      \For{$x\in V$}{
        $\psi[x]\assign 0$\;
        $\outside[x]\assign \infty$\;
        $\Adji[x]\assign \{\}$\;
      }
      \For{$1\leq i \leq m$,\text{ index of a hyperarc }
        $(T_{i}=\{x_{1},\ldots,x_{k}\}) \rightarrow^{l_i} \{h_{i}\}$}{
        \lFor{$1\leq j \leq k$}{$\Adji[h_i] \assign \Adji[h_i] \union \{x_j\} $\;}
      }
      $\outside[y]\assign 0$\;
      $Q \assign \PQ{CREATE}()$\;
      $\PQ{INSERT}(Q,y,0)$\;
      \While{$Q \neq \emptyset$}{
        $x\assign \PQ{EXTRACT-MIN}(Q)$\;
        \For{$i \in \Adji[x]$}{
                              \tcc{ edge $i$ with $x$ as a head}
          $c\assign \outside[x]+l_i+\sum_{(t,m)\in T_i})m\inside[t]$ \tcc{ c=total cost of $X\leadsto e_i \leadsto y$}\;
          \For{$t\in T_i$}{
            $c' \assign c-\inside[t]$ \tcc{ $c'$ is the proposed improved outside cost for $t$ through $e_i$, removing $X\leadsto t$}\;
            \If{$c'<\alpha[t]$}{
              \lIf{$\outside[h_{i}] = \infty$}{$\PQ{INSERT}(Q,t,c')$\;}
              \lElse{$\PQ{DECREASE-KEY}(Q,t,c')$\;}
              $\psi[t] \assign i$\;
              $\outside[t] \assign c'$\;
            }
          }
        }
      }
    }

    \label{algo_dijkstra}
  \end{algorithm}

    \newcommand\goodenough{\kappa}
  \begin{algorithm}
    \DontPrintSemicolon

    \caption{
      Prune relatively-useless vertices and hyperarcs
    }

    \KwIn{

      $\inside[v]$ and $\outside[v]$, the Viterbi inside and outside costs of
      each vertex V over all hyperpath-trees from $X\leadsto y$ (computed with
      \algname{ViterbiInside} and \algname{ViterbiOutside}) in a hypergraph
      $G=(V,E)$ with $m$ hyperarcs $E=\{e_1,\ldots,e_m\}$ indexed by $1\leq
      i\leq m$.  Each hyperarc has \hastails and \hashead.  The cost for
      hyperpath-tree from $X\leadsto h_e$ using edge $e$ and the best
      hyperpath-trees from $X$ to each of its tails $t$ with cost $\inside[t]$
      is $c_{e}=l_{e}+\sum_{t\in T_{i}}m_{t}\inside[t]$, where $l_{e}$ is the
      weight on hyperarc $e$ and $m_{t}$ is a weight, e.g. the number of occurrences
      of $t$ in the rhs of a grammar production.

      $\delta$ is a beam (cost distance from the best hyperpath-tree).

    }

    \KwOut{

      For all $x\in{V\union E}$, $\gamma[x]$ is the cost of the best
      hyperpath-tree $t\in{(X\leadsto_{G}y)}$ such that $x$ is used in $t$, or
      $\infty$ if none exists, $\goodenough[x]=\true$ iff that cost is not more worse
      than $\delta$ from the best $\inside[y]$.

      Time complexity is $O(t)$ where $t$ is the total size of the input.
      (total complexity including \algname{ViterbiInside} is $O(n\lg{n}+t)$).

    }

    \Begin{
        \;
      $l \assign \inside[y] + \delta$\;
      \For{$v\in{V}$}{
        $\gamma[v] \assign \inside[v]+\outside[v]$\;
      }
      \For{$e\in{E}$}{
        $\gamma[e] \assign \outside[h_e]+l_{e}+\sum_{t\in T_{i}}m_{t}\inside[t]$\;
      }
      \lFor{$x\in{V\union E}$}{
        $\goodenough[x] \assign (\gamma[x] \leq l)$\;
      }
    }

    \label{algo_prune_relatively_useless}
  \end{algorithm}

  \bibliographystyle{fullname}
  \bibliography{tree}

\begin{thebibliography}{}

\bibitem[\protect\citename{Dijkstra}, 1959]{dijkstra}
Dijkstra, E.~W.
\newblock 1959.
\newblock A note on two problems in connection with graphs.
\newblock {\em Numerical Mathematics}, 1:269--271.

\bibitem[\protect\citename{Graehl and Knight}, 2004]{TTT}
Graehl, Jonathan and Kevin Knight.
\newblock 2004.
\newblock Training tree transducers.
\newblock In {\em Proceedings of the 2004 Meeting of the North American chapter
  of the Association for Computational Linguistics (NAACL-04)}.

\bibitem[\protect\citename{Hopcroft and Ullman}, 1979]{hopcroft}
Hopcroft, John and Jeffrey Ullman.
\newblock 1979.
\newblock {\em Introduction to Automata Theory, Languages, and Computation}.
\newblock Addison-Wesley Series in Computer Science. Addison-Wesley, London.

\bibitem[\protect\citename{Knight and Graehl}, 2005]{poweroftree}
Knight, K. and J.~Graehl.
\newblock 2005.
\newblock An overview of probablistic tree transducers for natural language
  processing.
\newblock In {\em Proceedings of the Sixth International Conference on
  Intelligent Text Processing and Computational Linguistics (CICLing)}.

\bibitem[\protect\citename{Knuth}, 1977]{knuthgrammar}
Knuth, D.
\newblock 1977.
\newblock A generalization of {Dijkstra's} algorithm.
\newblock {\em Info. Proc. Letters}, 6(1).

\bibitem[\protect\citename{Lari and Young}, 1990]{InsideOutside}
Lari, K. and S.~J. Young.
\newblock 1990.
\newblock The estimation of stochastic context-free grammars using the
  inside-outside algorithm.
\newblock {\em Computer Speech and Language, 4}, pages 35--56.

\end{thebibliography}

\end{document}